\documentclass[aps,prb,superscriptaddress,twocolumn,showpacs]{revtex4-2}
\usepackage{amssymb}
\usepackage{epsfig}
\DeclareMathAlphabet{\pazocal}{OMS}{zplm}{m}{n}            
\usepackage{graphicx}
\usepackage[usenames, dvipsnames]{xcolor}
\usepackage{bm}
\usepackage[normalem]{ulem}     
\usepackage{soul}
\usepackage{amsmath}
\usepackage{braket} 
\usepackage{rotating}
\usepackage{multirow} 
\usepackage{mhchem}
\DeclareMathAlphabet{\pazocal}{OMS}{zplm}{m}{n}            
\usepackage{graphicx}
\usepackage{hyperref}
\usepackage[normalem]{ulem}
\usepackage{braket} 
\usepackage{booktabs}
\usepackage{tabularx}
\usepackage{multirow}
\usepackage{graphicx}

\newcommand\redsout{\bgroup\markoverwith{\textcolor{red}{\rule[0.5ex]{2pt}{0.4pt}}}\ULon}
\newcommand\bluesout{\bgroup\markoverwith{\textcolor{blue}{\rule[0.5ex]{2pt}{0.4pt}}}\ULon}

\newcommand{\SPhide}[1]{{}}

\begin{document}
\title{Why Is the $d$-Wave Spin Splitting in CuF$_2$ Bulk-Like?}        
\author{Muskan} 
\affiliation{Department of Physics, Indian Institute of Technology Bombay, Mumbai 400076, India}

\author{Subhadeep Bandyopadhyay}  
\affiliation{Consiglio Nazionale delle Ricerche (CNR-SPIN),  Unità di Ricerca presso Terzo di Chieti, c/o Università G. D'Annunzio, 66100 Chieti, Italy}

\author{Sayantika Bhowal}
\email{sbhowal@iitb.ac.in}
\affiliation{Department of Physics, Indian Institute of Technology Bombay, Mumbai 400076, India}

\date{\today}

\begin{abstract}
{With the advent of non-relativistic spin splitting in collinear compensated antiferromagnets, several candidate materials have also been proposed, among which the family of transition-metal difluorides stands out as a prominent example. Within this family, most members exhibit planar $d$-wave spin splitting, whereas CuF$_2$ shows bulk $d$-wave splitting with an explicit $k_z$ dependence. In this work, we show that this transition from planar to bulk $d$-wave splitting in CuF$_2$ is primarily driven by the antipolar displacements of the F ions, which are absent in the tetragonal rutile structure of the other family members. Our calculations reveal that these additional structural distortions introduce an extra plane of anisotropic magnetization density, giving rise to an additional totally symmetric component of the magnetic octupole tensor. The $k$-space representation of this octupole component, consequently, dictates an additional direction of spin splitting, thereby transforming the $d$-wave spin splitting pattern from planar to bulk-like. We further analyze the effect of spin–orbit coupling on the magnetic octupoles and the resulting spin splitting in the band structure. Our work highlights the possibility of controlling the pattern of non-relativistic spin splitting through structural modifications, for example, via the application of external pressure.
}

\end{abstract}

\maketitle

\section{introduction}

Non-relativistic spin splitting (NRSS) of spin-polarized bands in collinear compensated antiferromagnets with certain symmetries, in the absence of both spin–orbit coupling (SOC) and net magnetization, has recently emerged as one of the central topics of interest in condensed matter research \cite{Hayami2019, Yuan2020, Yuan2021, Smejkal2022PRX, YuanZunger2023, Guo2023, Zeng2024, Lee2024PRL, Krempask2024, Reimers2024, Aoyama2024, Lin2024, Kyo-Hoon2019, Libor2020, Libor2022Review, Paul2024, Bai2024, Hayami2020PRB, Jungwirth2024, Cheong2024, Radaelli2025, Bhowal2025}. The discovery of NRSS provides an alternative route to achieving spin splitting beyond well-known mechanisms such as exchange-driven splitting in ferromagnets or SOC-induced Rashba-like splitting \cite{Rashba1960, Bychkov1984} in noncentrosymmetric systems. In contrast to these effects, which are central to conventional spintronics applications \cite{Manchon2015}, the emergence of large NRSS in collinear antiferromagnets without SOC and while maintaining zero net magnetization opens new avenues in antiferromagnetic (AFM) spintronics and related fields. These include the spontaneous anomalous Hall effect \cite{Libor2020, Libor2022NatRev, Helena2020, Feng2020, Betancourt2021, SmejkalAHE2022, Cheong2024, Sato2024}, unconventional spin-transport phenomena \cite{Naka2019, Hernandez2021, Shao2021, Bose2022, Bai2022, Karube2021, Hu2024}, giant magnetoresistance \cite{Libor2022}, superconductivity \cite{Mazin2022, Zhu2023, Banerjee2024, Chakraborty2024, Zhang2024, Lee2024}, and chiral magnons \cite{Libor2023, McClarty2024, Liu2024, Morano2024, Bandyopadhyay2024}.

Since its discovery, numerous AFM materials exhibiting NRSS in their band structures have been identified \cite{Yuan2021, Smejkal2022PRX, Guo2023, Zeng2024, Hayami2020PRB, Cheong2025, Lee2024, Dale2024, Reimers2024, Yang2025}. Among these, the family of transition-metal difluorides MF$_2$ (M = Mn, Co, Fe, Ni, Cu, etc.) is particularly notable \cite{Yuan2020, Smejkal2022PRX, BhowalSpaldin2024}. Interestingly, many members of this family have been extensively studied for decades \cite{Stout1942, Erickson1953, Dzialoshinskii1958, Borovikromanov1960, Baruchel1980, Baruchel1988}, yet NRSS was identified only recently due to advances in understanding the underlying symmetry requirements. Moreover, recent theoretical developments \cite{Hayami2019, BhowalSpaldin2024, Verbeek2024, Nag2024} have revealed that NRSS is intimately connected to the ferroically ordered magnetic multipoles, providing a framework for controlling both the magnitude and sign of NRSS \cite{Stroppa_2025, Bandyopadhyay2025, Ray2025, Bandyopadhyay2024, Martinelli2025}.

\begin{figure}[t]
    \centering
    \includegraphics[width=\columnwidth]{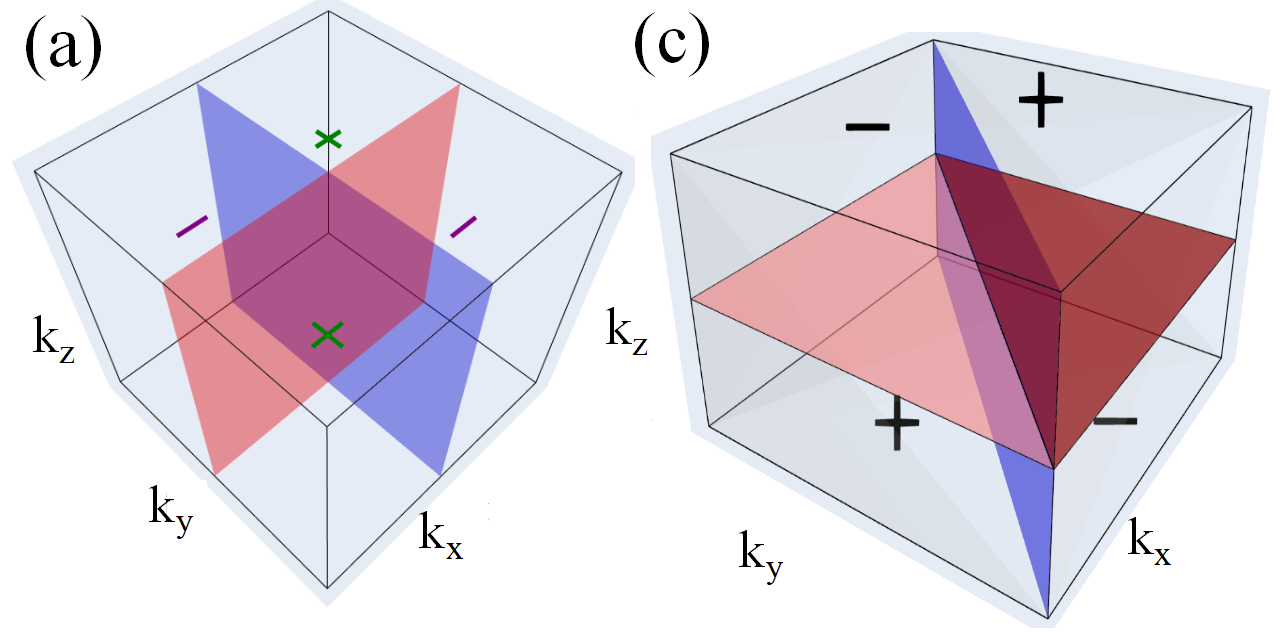}\\
    \vspace{2mm} 
    \includegraphics[width=\columnwidth]{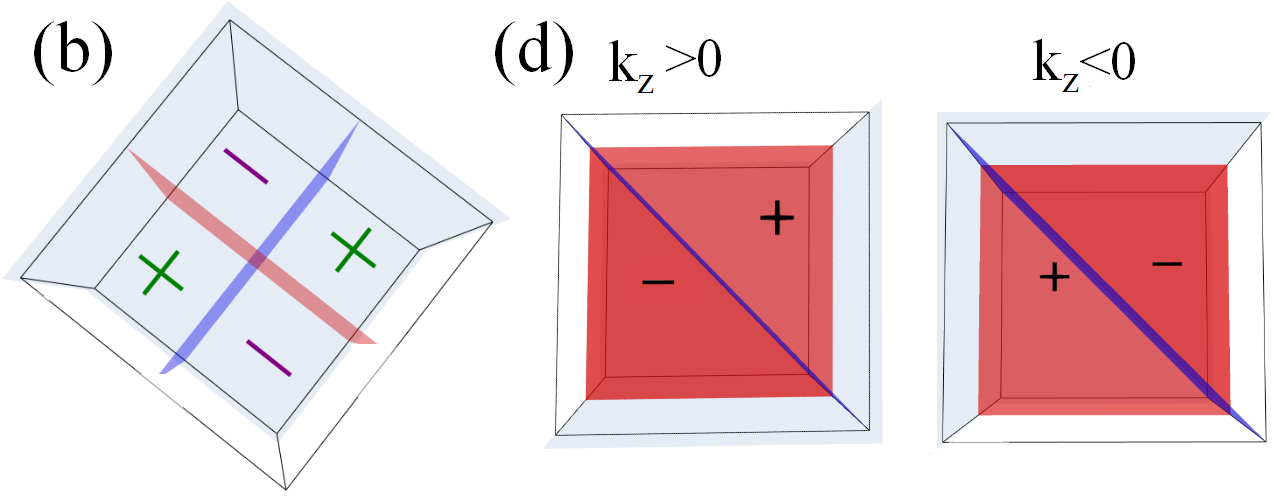}
    \caption{ Schematic illustrations of planar vs. bulk $d$-wave spin splitting. The (a) side and (b) top views of the planar $d$-wave spin splitting across different parts of the BZ. The same (c) side and (d) top views of the bulk-$d$ wave spin splitting. The red and blue planes indicate the nodal planes of degenerate up ($E_{\uparrow}$)- and down-spin ($E_{\downarrow}$) polarized bands. The $\pm$ sign shows the sign of the energy splitting, $\Delta E = E_{\uparrow} - E_{\downarrow}$. The reversal of the NRSS energy, as shown in (d), for $k_z > 0$ ({\it left}) and $k_z < 0$ ({\it right}) depicts the $k_z$ dependence of the bulk $d$-wave splitting.  }
    \label{fig1}
\end{figure}

Several members of the MF$_2$ family (M = Mn, Co, Fe, and Ni) crystallize in the rutile structure \cite{StoutReed1954, Strempfer2004, Hariki2025, KURKCU201617} and exhibit planar $d$-wave spin splitting characterized by two nodal planes \cite{SmejkalJairo2022, BhowalSpaldin2024}. As illustrated schematically in Figs.~\ref{fig1}a and \ref{fig1}b, the presence of two nodal planes, along which the up- and down-spin bands remain degenerate, causes the spin-splitting energy to change sign across different regions of the Brillouin zone (BZ), resulting in a $d$-wave symmetry. We note that the $d$-wave spin-splitting pattern shown in Fig.~\ref{fig1}b is independent of $k_z$ and is therefore referred to as planar $d$-wave splitting. This behavior has been linked to the ferroically ordered magnetic octupoles, which describe the anisotropic magnetization density in these materials.

In contrast, CuF$_2$ crystallizes in the monoclinic $P2_1/c$ structure \cite{TAYLOR1974257, reinhardt1999, fischer1974} and has been proposed as a candidate for bulk $d$-wave spin splitting \cite{Smejkal2022PRX}. In this case, one of the nodal planes lies within the three-dimensional BZ, as shown in Fig.~\ref{fig1}c. Consequently, the spin splitting becomes $k_z$ dependent, with the spin-splitting energy changing sign as a function of $k_z$ (see Fig.~\ref{fig1}d), a phenomenon referred to as bulk $d$-wave splitting. This raises the key issue of why CuF$_2$ exhibits bulk $d$-wave splitting, whereas other members of the MF$_2$ family exhibit planar $d$-wave splitting.

In this work, we address this question using the multipolar framework combined with phonon-based structural analysis. We show that the octupolar framework developed for planar $d$-wave splitting can be extended to describe bulk-like $d$-wave splitting. In the latter case, two distinct totally symmetric components of the magnetic octupole tensor exist, in contrast to the single component present in AFMs with planar $d$-wave NRSS. We demonstrate that the emergence of this second octupole component is directly linked to additional structural distortions in the lower-symmetry CuF$_2$, compared to AFMs exhibiting planar $d$-wave NRSS within the same family. These distortions generate anisotropic magnetization density in an additional plane, which is captured by the new totally symmetric component of the magnetic octupole moment. Consequently, the representation of this component in $k$ space dictates the direction of NRSS and modifies the splitting pattern from planar to bulk-like.

Our results highlight the crucial role of structural distortions, quantified through phonon modes, in determining the NRSS pattern in CuF$_2$. Furthermore, the implications of this study extend beyond CuF$_2$, suggesting routes to tune NRSS through controlled structural distortions, for example, by applying external pressure \cite{Cihan_CoF2_2016, Barreda_CoF2_2013}. These insights deepen our understanding of the interplay between structure and NRSS and provide a pathway toward designing controllable, spin–orbit-free spintronic devices.

The remainder of this manuscript is organized as follows. Section~\ref{sec2} describes the computational methodology and structural details of CuF$_2$. Section~\ref{sec3} presents the results and discussion, beginning with an analysis of the electronic structure of CuF$_2$. This is followed by a comprehensive investigation of magnetic multipoles and NRSS in both the ground-state monoclinic structure and a hypothetical rutile tetragonal structure constructed to mimic other members of the MF$_2$ family. We then analyze the structural distortion modes connecting the hypothetical tetragonal $P4_2/mnm$ structure to the ground-state $P2_1/c$ structure of CuF$_2$. The effect of spin–orbit coupling on the multipoles and band structure, particularly on NRSS, is also discussed. Finally, we summarize our findings and outline their implications for the rapidly growing field of NRSS.

\section{Crystal Structure and Computational Methods}\label{sec2}

\begin{figure}[!t]
    \includegraphics[width=\columnwidth]{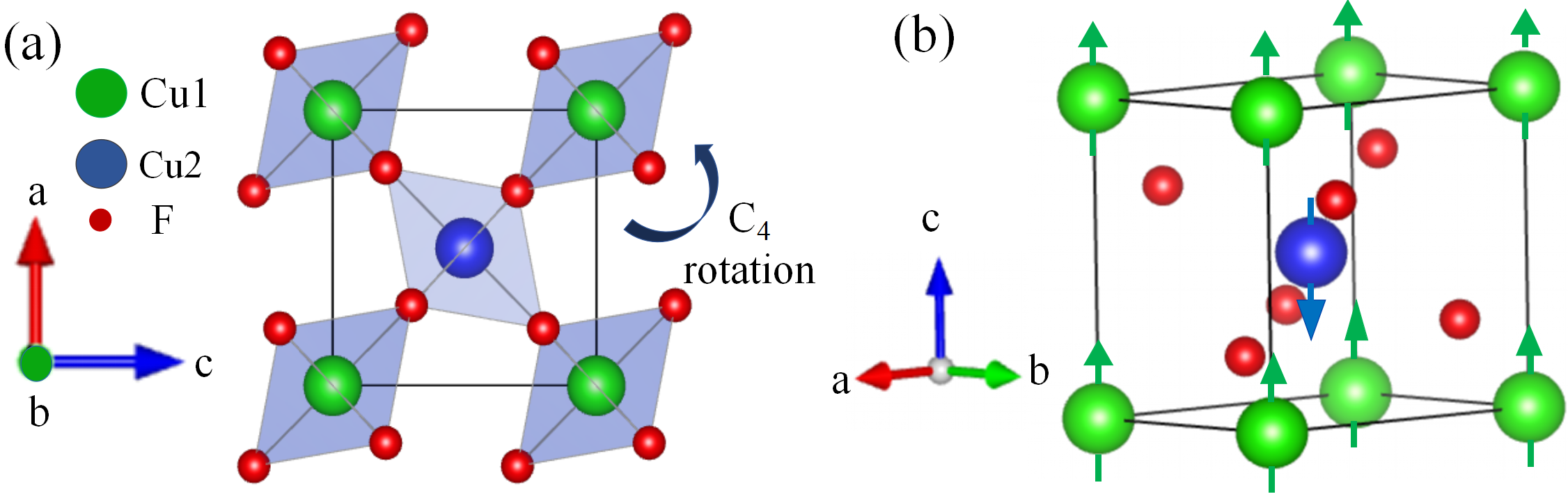}
    \caption{ (a) Crystal Structure of $\mathrm{CuF}_2$. The Cu sublattices, viz., Cu1 and Cu2, are indicated in different colors. The CuF$_6$ octahedra around the Cu1 and Cu2 are rotated by 90\textdegree{} with respect to each other. (b) The AFM configuration of CuF$_2$. }
    \label{fig2}
    \vspace{-0.5em}
\end{figure}

$\mathrm{CuF}_2$ crystallizes in the monoclinic structure with space group $P2_1/c$ (No. 14), derived from a distorted rutile-type framework~\cite{Taylor1974}. As shown in Fig. \ref{fig2}, the unit cell of $\mathrm{CuF}_2$ contains two formula units, i.e.,  two Cu atoms and four F atoms. Each Cu atom is octahedrally coordinated by six F atoms. As depicted in Fig. \ref{fig2}a, the CuF$_6$ octahedra around the two Cu atoms are rotated by 90$^\circ$ relative to each other, which plays a crucial role in the NRSS of CuF$_2$. To understand the bulk $d$-wave splitting in $\mathrm{CuF}_2$ in contrast to the planar $d$-wave splitting in the rutile structure of other members of the transition metal di-fluoride ($M$F$_2$) family, we also constructed and analyzed the hypothetical $P4_2/mnm$ structure of CuF$_2$.

All electronic structure calculations of CuF$_2$, presented in the work, are carried out using the projector augmented-wave (PAW) method ~\cite{Blochl1994} as implemented in the Vienna \emph{Ab initio} Simulation Package (VASP)~\cite{KresseHafner1993}. The exchange-correlation functional is treated within the generalized gradient approximation (GGA) with the following valence-electron configurations: Cu [Ar] $3d^{10} 4s^1$ and F: [He]$2s^22p^5$. 
To achieve self-consistency, we have used a plane-wave energy cutoff of 700\,eV and the Brillouin zone is sampled with a \(10 \times 14 \times 10\) \emph{k}-point mesh. To take into account the correlation effects, an effective Hubbard interaction of $U_{\rm eff}=U-J=$ 4 eV, is added to the Cu-$d$ states~\cite{Dudarev1998}. Calculations are performed both in the presence and absence of the spin-orbit interaction. 

The phonon calculations are carried out
using the finite displacement method as implemented in the PHONOPY code~\cite{Phonopy}. A \(2 \times 2 \times 2\) supercell of the high-symmetry hypothetical $P4_2/mnm$ structure of CuF$_2$ is used to compute the phonon band structure. All phonon calculations are performed using the relaxed crystal structure of CuF$_2$. The structural relaxations are performed until the Hellmann–Feynman forces on all atoms are reduced below 0.005\,eV/\AA.

The atomic-site charge and magnetic multipoles are computed within an atomic sphere around the Cu atom \cite{Spaldin2013} by decomposing the DFT computed density matrix $\rho_{lm,l'm'}$ into irreducible (IR) spherical tensor components $w_t^{kpr}$~\cite{Spaldin2013}. In this notation, $k$, $p$, and $r$ correspond to the spatial index, spin index, and the rank of the tensor, respectively, while $t$ denotes the specific tensor component. The rank of the tensor, $r$, ranges from $|k - p|$ to $k + p$, and for a specific $r$, $t$ ranges from $-r$ to $r$. 

The two multipoles that are at the center of the current work are the charge quadrupole and the magnetic octupole. The corresponding IR spherical tensor components are $w^{202}$ and $w^{21r}$ respectively, indicating the same spatial index $k=2$ for both multipoles, while their spin dependences are different. The former being a charge multipole, preserves the time-reversal ($\cal T$) symmetry while the latter breaks $\cal T$ symmetry, leading to $p=0$ and 1, respectively. Consequently, the rank $r=0$ for the charge quadrupole moment, while for the magnetic octupole, there are three possible values of $r$, viz., $r = 1, 2, 3$. The different values of $r$ correspond to different components of the octupole tensor. For example, $w^{211}$ with 3 components ($t=-1, 0, 1$) constitutes the nonsymmetric component of the octupole tensor, known as moment of toroidal moment $t_i^{(\tau)}$; another non-symmetric component, $w^{212}$ with 5 components ($t=-2,..., 2$), called the quadrupole moment of toroidal moment, $Q_{t}^{(\tau)}$; and $w^{213}$, corresponds to the totally symmetric part of the magnetic octupole tensor, and is represented by $\mathcal{O}_{3m}$, with $m = -3, ..., +3$ denoting its seven components~\cite{Urru2022}.

\section{Results and Discussions}\label{sec3}

In this section, we present our results, describing the origin of the bulk $d$-wave spin splitting in CuF$_2$. In addition to the ground state monoclinic structure,  we also consider a hypothetical tetragonal structure of CuF$_2$ to investigate the evolution of the NRSS from the rutile structure, adopted by members of the $M$F$_2$ family with planar $d$-wave splitting, to the monoclinic structure. We first demonstrate that CuF$_2$ exhibits bulk $d$-wave spin splitting in its band structure, and discuss the multipolar framework to understand the obtained bulk $d$-wave spin splitting. Finally, we identify the structural distortions that lead to changes in the magnetisation density and, hence, the transformation of  
$d$-wave spin splitting from planar to bulk-like.

\subsection{Basic electronic structure}

\begin{figure}[!t]
\includegraphics[width=1.0\columnwidth]{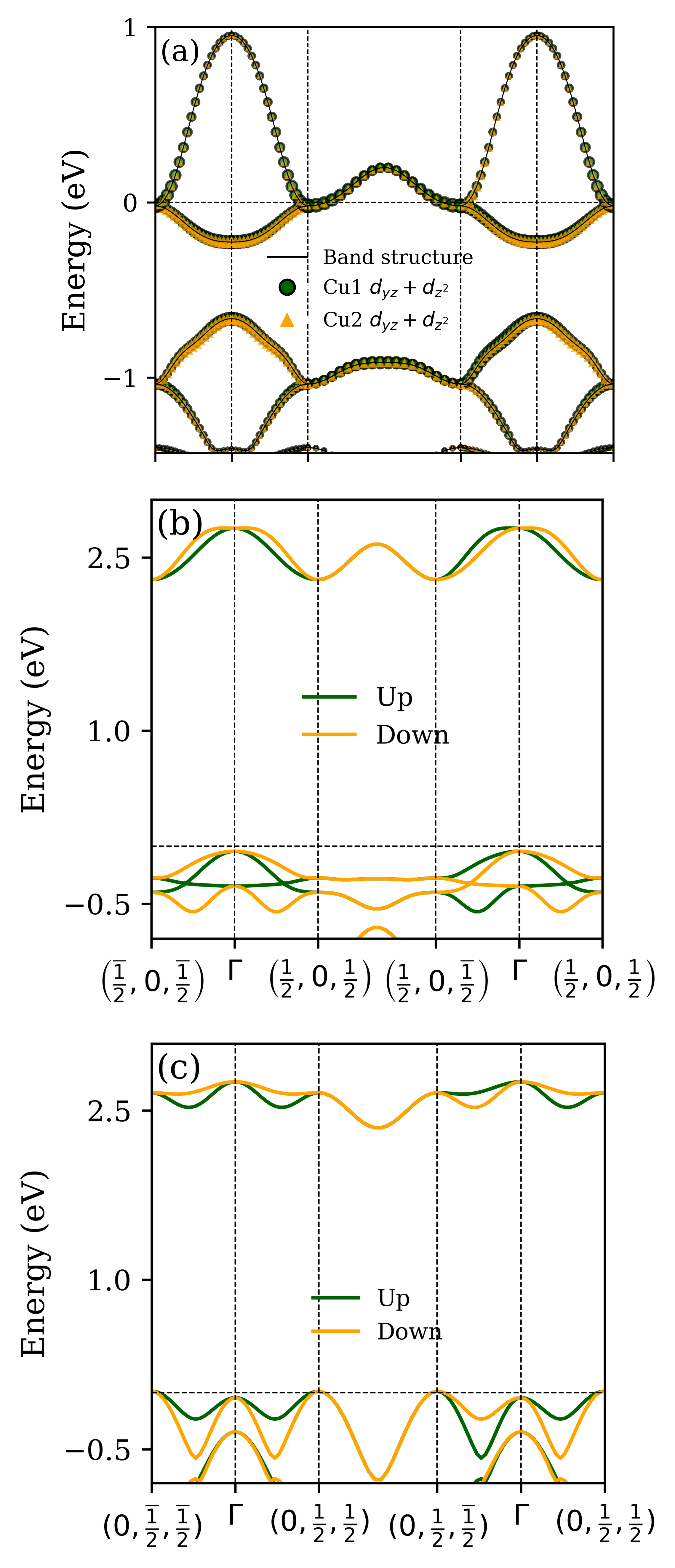}
    \caption{Band structures of \( \mathrm{CuF_2} \), computed for the (a) non-magnetic configuration and (b-c) the AFM configuration (see Fig. \ref{fig2}b). NRSS along (b) \( k_x \)-\( k_z \), and (c)  \( k_y \)-\( k_z \) directions are apparent from the band structures in the presence of antiferromagnetism. }
\label{fig3}
    \vspace{-0.5em}
\end{figure}

We begin by analyzing the basic electronic structure of CuF$_2$. 
To start with, we compute and analyze the nonmagnetic electronic structure of CuF$_2$. The computed band structure of CuF$_2$ is shown in Fig. \ref{fig3}a. Within the nonmagnetic configuration, the system is metallic, as evident from the plot of the band structure. The metallic behavior is consistent with the partially filled Cu-3$d$ orbitals (Cu$^{2+} : 3d^9$) in \( \mathrm{CuF}_2 \). A strong hybridization between Cu-$d$ and F-$p$ orbitals is also evident from the plot of partial densities of states (DOS) (see Appendix~\ref{app1}). 

We further analyze the orbital character of the bands across the Fermi energy within the nonmagnetic configuration, and this is shown in Fig. \ref{fig3}a. As evident from the figure, the pair of bands near the Fermi energy has a predominant Cu-\( d_{yz} \) and Cu-\( d_{z^2} \) orbital character. More interestingly, however, the top band has dominant contribution from Cu1 (corner Cu atom) sublattice while the bottom band is predominantly of Cu2 sublattice (central Cu atom) character along $(\frac{\bar1}{2},0,\frac{\bar1}{2}) -\Gamma-(\frac{1}{2},0,\frac{1}{2}) $ path. 
This leads to a splitting between nonmagnetic bands of two opposite sublattice character.  

After analyzing the nonmagnetic electronic structure of CuF$_2$, we investigate the effect of magnetism within the DFT+$U$ framework. For this, we perform calculations for collinear ferromagnetic (FM) and AFM configurations, in which the two Cu ions in the unit cell have parallel and antiparallel spins, respectively. The results of our calculations are shown in Table \ref{tab1}. From our computed total energies (see Table \ref{tab1}), we find that the AFM configuration is lower in energy compared to the FM configuration, suggesting an AFM ground state of CuF$_2$.

The computed band structure for the AFM configuration is shown in Fig. \ref{fig3}b. We note that, in contrast to the nonmagnetic case, the band structure depicts an insulating behavior within the AFM+$U$ framework, consistent with the insulating behavior observed in experiments \cite{OlaldeVelasco2011}. This also highlights the effect of Hubbard $U$ in driving the insulating property of CuF$_2$. We further note the presence of spin splitting between up and down spin-polarized bands of CuF$_2$. Importantly, this spin splitting occurs even without the SOC, and, hence, is of non-relativistic origin. We discuss the NRSS in more detail in section \ref{NRSS}.

\begin{table}[h]
  \centering
  \caption{Comparison of the computed total energies and spin magnetic moments for the FM and AFM configurations of $\mathrm{CuF}_2$.}
  \label{tab1}
  \resizebox{0.8\columnwidth}{!}{%
    \begin{tabular}{|c|c|c|c|}
      \hline
      Config. & {$\Delta E$} & {Moment} & {Total Moment} \\
              & {(meV/f.u.)} & {($\mu_B$/Cu)} & {($\mu_B$/f.u.)} \\
      \hline
      FM  & 0    & 0.81 & 2.00 \\
      \hline
      AFM & -24  & 0.78 & 0.00 \\
      \hline
    \end{tabular}%
  }
\end{table}

\subsection{Multipole Analysis}\label{Multipole Analysis}

The AFM configuration (see Fig. \ref{fig2}b) of CuF$_2$ breaks the {\it global} $\cal T$ symmetry, i.e., time-reversal combined with translation is not a symmetry of the system due to the rotated CuF$_6$ octahedral units, as discussed earlier. The breaking of  {\it global} $\cal T$ symmetry implies the presence of a ferroically ordered magnetic multipole. Since the magnetic dipole has an antiferroic order in CuF$_2$, this suggests there must be some higher-order magnetic multipole that has a ferroic ordering. Indeed, a ferroic ordering of rank-3 magnetic octupoles has been identified in the AFM tetragonal transition-metal difluorides \cite{BhowalSpaldin2024}. 

Motivated by this, we perform the multipole calculations and analyze all the charge and magnetic multipoles present in CuF$_2$. Table \ref{tab2} summarizes the computed charge and magnetic multipoles, relevant to our current discussion. As listed in Table \ref{tab2}, the monoclinic structure of CuF$_2$ hosts two charge quadrupole moment components \( \mathcal{Q}_{{1}^{}} \) and \( \mathcal{Q}_{1^-} \) with antiferroic order at the Cu ions. These two charge quadrupole moments represent anisotropic charge density on the $xz$ and $yz$ planes of CuF$_2$, respectively. Interestingly, the antiferroic order of these charge quadrupole components is identical to that of the antiferroic order of the spin dipole moment of the Cu ions (see Table \ref{tab2}). This, consequently, leads to the ferroic order of several magnetic octupole components, ${\cal O}_{ijk}=\int r_i r_j \mu_k(\vec{r})d^3r$ \cite{BhowalSpaldin2024}, as listed in Table \ref{tab2}. 

The magnetic octupole is a rank-3, inversion symmetric but $\cal T$-broken magnetic multipole, the components of which describe the anisotropy in the magnetization along specific planes of the crystal structure. Among the ferroically ordered magnetic octupoles in CuF$_2$, the components ${\mathcal O}_{31}$, ${Q}^{(\tau)}_{1^-}$ and $\tau^{(\tau)}_x$  describe an anisotropic magnetization density on the $xz$ plane with the spin-polarization along $\hat z$, represented in real space as $xzm_z$. On the other hand, octupole components ${\mathcal O}_{31^-}$, ${Q}^{(\tau)}_{1}$ and $\tau^{(\tau)}_y$, have the real-space representation $yzm_z$, indicating the presence of anisotropic magnetization density on the $yz$ plane with the spin-polarization along $\hat z$.

\begin{table}[h]
\centering
\large
\renewcommand{\arraystretch}{1.2}
\caption{Relevant multipoles at the Cu ions in the ground state monoclinic structure and artificial tetragonal structure of $\mathrm{CuF}_2$ for the AFM configuration. 
Cu1 and Cu2 denote the corner and central Cu atoms in the unit cell (see Fig.~\ref{fig2}a). 
The $\pm$ signs indicate the relative orientation of each atomic-site multipole between Cu1 and Cu2.}
\label{tab2}

\resizebox{0.8\columnwidth}{!}{%
\begin{tabular}{|c|c|c|c|c|}
\hline
\multicolumn{5}{|c|}{\textbf{Monoclinic}} \\
\hline
\textbf{Multipole} & \textbf{$W^{kpr}$} & $t$ & \textbf{Component} & \textbf{Cu1 / Cu2} \\
\hline

\multirow{2}{*}{Charge Quadrupole} 
  & \multirow{2}{*}{$W^{202}$} 
    & $-1$ & $\mathcal{Q}_{1^-}$ & $- / +$ \\
  & & $1$ & $\mathcal{Q}_{1}$ & $+ / -$ \\
\hline

Magnetic Dipole 
  & $W^{011}$ 
    & $0$ & $m_z$ & $- / +$ \\
\hline

\multirow{7}{*}{Magnetic Octupole}
  & \multirow{2}{*}{$W^{213}$}
    & $-1$ & $\mathcal{O}_{31^-}$ & $- / -$ \\
  & & $1$ & $\mathcal{O}_{31}$ & $+ / +$ \\
\cline{2-5}

  & \multirow{2}{*}{$W^{212}$}
    & $-1$ & $Q^{(\tau)}_{1^-}$ & $+ / +$ \\
  
  & & $1$ & $Q^{(\tau)}_{1}$ & $+ / +$ \\
\cline{2-5}

  & \multirow{2}{*}{$W^{211}$}
    & $-1$ & $\tau^{(\tau)}_{y}$ & $- / -$ \\
  & & $1$ & $\tau^{(\tau)}_{x}$ & $+ / +$ \\
\hline

\multicolumn{5}{|c|}{\textbf{Tetragonal}} \\
\hline
\textbf{Multipole} & \textbf{$W^{kpr}$} & $t$ & \textbf{Component} & \textbf{Cu1 / Cu2} \\
\hline

Charge Quadrupole
  & $W^{202}$ 
     & $1$ & $\mathcal{Q}_{1}$ & $+ / -$ \\
\hline
Magnetic Dipole
  & $W^{011}$ 
    & $0$ & $m_z$ & $- / +$ \\
\hline

\multirow{4}{*}{Magnetic Octupole}
  & $W^{213}$
    & $1$ & $\mathcal{O}_{31}$ & $+ / +$ \\
\cline{2-5}

  & {$W^{212}$}
    & $-1$ & $Q^{(\tau)}_{1^-}$ & $+ / +$ \\
  
\cline{2-5}

  & $W^{211}$
    & $1$ & $\tau^{(\tau)}_{x}$ & $+ / +$ \\
\hline

\end{tabular}
}
\end{table}

We further compute the multipoles for CuF$_2$ in its hypothetical tetragonal structure. 
In contrast to the ground state structure, as discussed above, our analysis reveals the presence of an antiferroically ordered charge quadrupole component $\mathcal{Q}_{1}$, along with an antiferromagnetically aligned magnetic dipole moment $m_z$ on the two Cu sublattices in the tetragonal structure of \( \mathrm{CuF_2} \). Consequently, in this case, we have ferroic ordering of the magnetic octupole components $\mathcal{O}_{31}$, ${Q}^{(\tau)}_{1^-}$, and $\tau^{(\tau)}_x$, describing an anisotropic magnetic density on the $xz$ plane with spin polarization along $\hat z$.

\subsection{Non-Relativistic Spin-Splitting} \label{NRSS}

We now discuss the consequences of the computed magnetic octupoles and the associated anisotropy in the magnetization density on the band structure of CuF$_2$, both in its ground state monoclinic structure and in the hypothetical tetragonal structure. 

To analyze the effect of magnetic octupoles on the band structure, we first examine their representations in $ k$-space. For all inversion symmetric multipoles, the corresponding $k$-space representations can be obtained directly from their real space representation by replacing $\vec r \rightarrow \vec k$. Following this and considering the spin-polarization of Cu along $\hat z$, we find the $k$-space representations of \( \mathcal{O}_{31^-} \) and \( \mathcal{O}_{31} \) are respectively \( \mu_z k_y k_z \) and \( \mu_z k_x k_z \). Physically, these imply the existence of energy splitting between bands of $\mu_z$ spin-polarization along any direction in the BZ of monoclinic CuF$_2$ with ($k_x \ne 0$, $k_z \ne 0$); and ($k_y \ne 0$, $k_z \ne 0$), respectively.

Consistent with the $k$-space representation, as shown in Figs.~\ref{fig3}b and c, we find an energy splitting along the \( k_x \)-\( k_z \) and \( k_y \)-\( k_z \) directions of the BZ of monoclinic CuF$_2$. 
As seen from Fig.~\ref {fig3}, the spin splitting energy depends on $k_z$. In particular, a reversal in the sign of the splitting energy takes place as the direction $\hat k_z$ is reversed, a characteristic of the bulk $d$-wave splitting.

Further, we have analyzed the band structure of the hypothetical tetragonal structure of \( \mathrm{CuF_2} \). In contrast to the monoclinic structure of \( \mathrm{CuF_2} \), our computed band structure shows that the spin splitting in the tetragonal structure is present only along the \( k_x\!-\!k_z \) direction, with no splitting along the \( k_y\!-\!k_z \) direction, leading to planar \( d \)-wave splitting. This is consistent with the presence of only one totally symmetric component of the ferroically ordered magnetic octupole, ${\cal O}_{31}$, in the tetragonal structure, the $k$-space representation of which is $k_xk_zm_z$. We note that this is consistent with other members of the \( \mathrm{MF_2} \) family, e.g., \( \mathrm{MnF_2} \), \( \mathrm{CoF_2} \), and \( \mathrm{FeF_2} \), that crystallize in the tetragonal rutile structure and exhibit the planar $d$-wave spin splitting \footnote{We note that the reported tetragonal structure of MnF$_2$ and other related materials have a different crystallographic orientation than the the tetragonal structure of \( \mathrm{CuF_2} \), discussed in the present work. This results in a different ferroic component of the magnetic octupole moment (e.g., \( \mathcal{O}_{32^{-}} \)) and direction ($k_x\!-\!k_y$) of NRSS in the previous reports. These are, however, physically consistent with our findings.}

\subsection{Structural distortion and phonon analysis}\label{Structural distortion}

As follows from the previous discussion, the pattern of the NRSS evolves from planar to bulk-like $d$-wave spin splitting as we go from the tetragonal to the monoclinic structure. To understand explicitly how the structural distortion affects the NRSS, here in this section we investigate the phonon distortion modes and their influence on the magnetic octupoles and consequently, on the NRSS.

\begin{figure}[h!]
\includegraphics[width=\columnwidth]{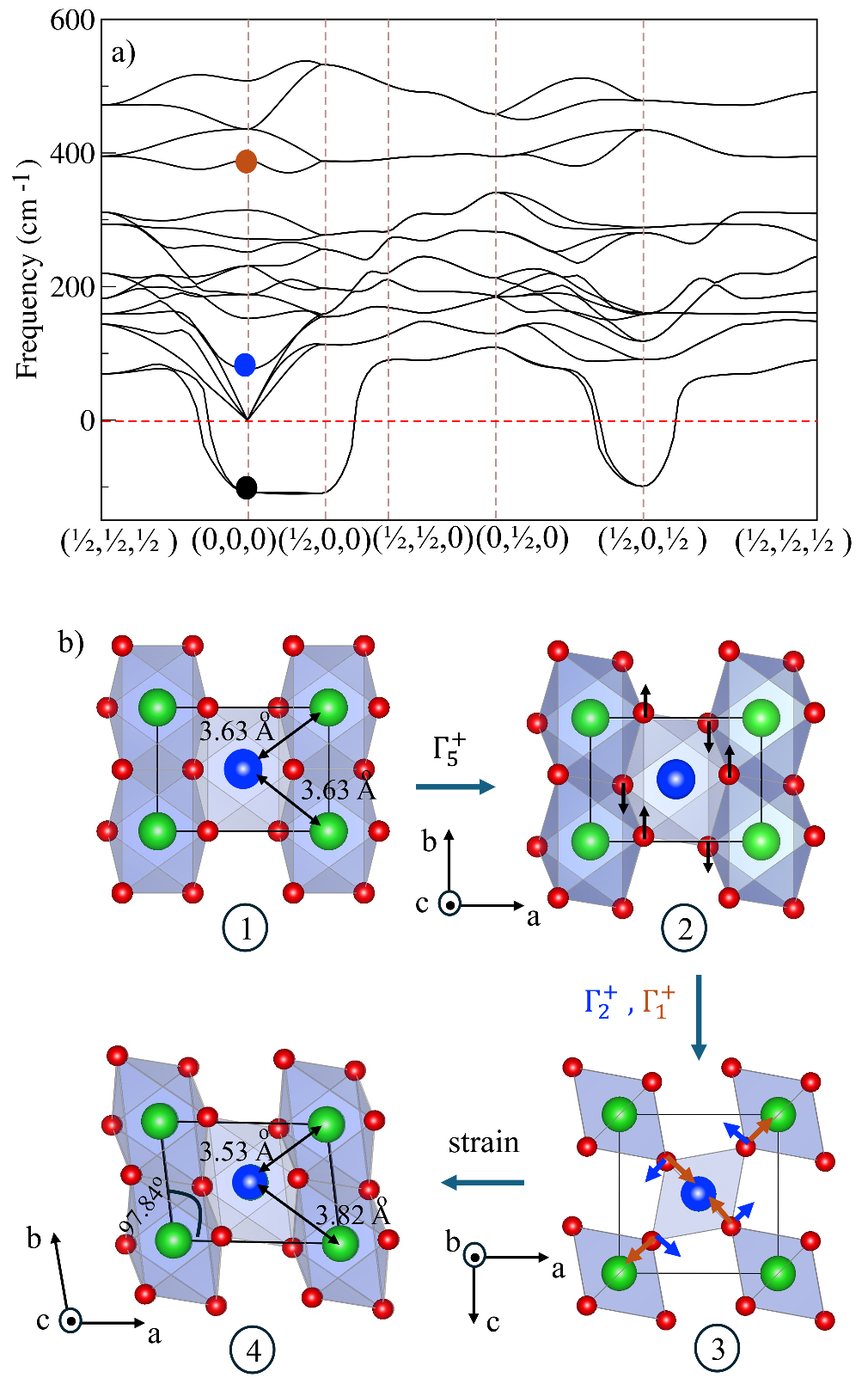}
 \caption{(a) Phonon band structure of CuF$_2$ in its hypothetical $P4_2/mnm$ structure.   Unstable phonon modes, marked with a black circle at the $\Gamma$-point, give rise to the antipolar displacements of the F atoms ($\Gamma_5^+$). Stable phonon modes, marked with blue and brown circles, are linked to the $\Gamma_2^+$ and $\Gamma_1^+$ distortions, respectively. (b) Structural change from tetragonal $P4_2/mnm$ (structure: 1) to the ground state monoclinic $P2_1/c$ (structure: 4). Structure: 2 is the intermediate  $P2_1/c$ structure, which includes only the atomic distortion of the unstable $\Gamma$-phonon modes, i.e., $\Gamma_5^+$  atomic distortion.
 Upon optimizing the ionic positions of structure:2, we obtain
 structure:3, which includes additional atomic distortions $\Gamma_2^+$ and $\Gamma_1^+$. Atomic displacements of the  $\Gamma_5^+$, $\Gamma_2^+$, and $\Gamma_1^+$ are represented by the black, blue, and brown arrows. 
 Upon optimizing the lattice parameters, we obtain the final ground state $P2_1/c$ structure (structure: 4), which includes all the atomic distortions and modification of the lattice parameters. 
 The change of the lattice angle $\gamma$ to 97.84$^\circ$ gives rise to a shearing strain in the $ab$ plane, producing two inequivalent Cu-Cu bonds with bond lengths of 3.53 and 3.82 \AA.}
 \label{fig4}
 \end{figure}

 We begin by investigating the evolution of the structural distortion from the tetragonal CuF$_2$ to the monoclinic CuF$_2$ structure. For this, we perform the phonon calculations for the $P4_2/mnm$ structure keeping the same AFM ordering of the Cu spins as in the ground state (see Fig. \ref{fig2}a). The computed phonon band structure is shown in Fig. \ref{fig4}a. As evident from the figure, the tetragonal $P4_2/mnm$ structure of CuF$_2$ is dynamically unstable. In particular, we find two unstable phonon branches along the $\Gamma-$X path with nearly no dispersion. These unstable phonon modes are degenerate 
   at the $\Gamma$ point with frequency 92$i$ cm$^{-1}$, and they give rise to an antipolar motion of the inequivalent F atoms along the crystallographic $b$-direction, as shown in Fig.\ref{fig4}b. These two degenerate phonon modes belong to the $\Gamma_5^+$ IR, which are predominantly responsible for lowering the structural symmetry to monoclinic $P2_1/c$. Upon performing ionic relaxation of the 	$P2_1/c$ structure, two additional atomic distortions emerge having $\Gamma_2^+$ and $\Gamma_1^+$ IR. $\Gamma_2^+$ is associated to orthorhombic $Pnnm$ symmetry, which corresponds to in-phase rotation of the CuF$_6$ octahedra along the $b$-axis. On the other hand, $\Gamma_1^+$ does not break $P4_2/mnm$ symmetry, and gives rise to the Jahn-Teller-like distortion of the  CuF$_6$ networks. Atomic displacements associated to the  $\Gamma_2^+$ and $\Gamma_1^+$ distortions are shown in Fig.\ref{fig4}, which are linked to the stable $\Gamma$ point phonon modes lying at energies 77 and 393 cm$^{-1}$ respectively.
   Finally,  relaxation of the lattice parameters induces additional contribution of the $\eta_{\Gamma_5^+}$, $\eta_{\Gamma_2^+}$, and $\eta_{\Gamma_1^+}$ strains,  giving rise to the ground state $P2_1/c$ structure with non-orthogonal lattice vectors, as shown schematically in Fig. \ref{fig4}b. 

By controlling the amplitudes of distortion of these modes, we generate intermediate structures that connect the tetragonal (with $0\%$ distortion) to the monoclinic (with $100\%$ distortion) structure (see the end structures in Fig. \ref{fig4}b). We note that, ignoring the strain modes, it is also possible to connect the tetragonal structure to the monoclinic structure with orthogonal lattice vectors (structure 3 in Fig. \ref{fig4}b). Herein, we focus on the first path for completeness and discuss our findings across different structures in the first path. The second path, however, can also provide crucial insight into the underlying cause of the changes in the NRSS pattern, which we discuss later.

\begin{figure}[t]
    \includegraphics[width=\columnwidth]{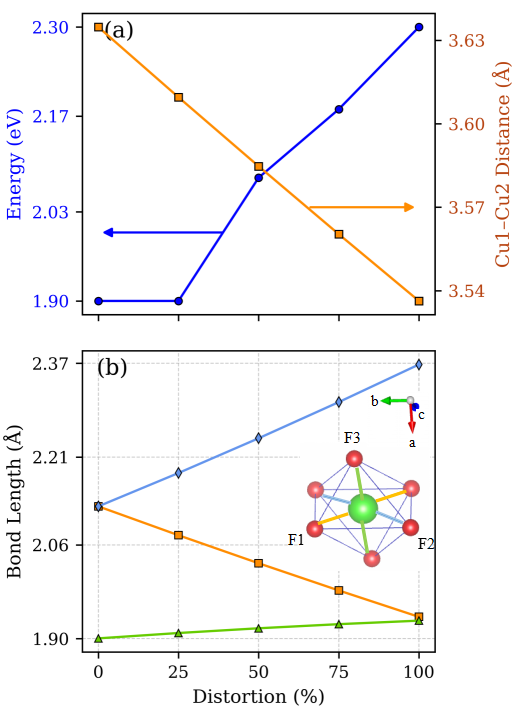}\\
    \vspace{2mm} 
    \caption{(a) The variation in Cu1–Cu2 distance and the electronic band gap as a function of amplitude of distortion (see text for details). (b) The changes in the Cu–F bond lengths as a function of distortion amplitude. The inset shows the CuF$_6$ octahedron, indicating different Cu-F bonds, color-coded with their corresponding variations shown in (b).} 
    \label{fig5}
\end{figure}

The change in the lattice angle ($\gamma$) to 97.84$^\circ$ in the structural evolution corresponding to the first path induces a shearing strain in the $ab$-plane, creating inequivalent Cu-Cu bonds in contrast to the tetragonal structure. 
As shown in Fig.~\ref{fig5}(a), the distance between the central and the corner Cu atoms (see Fig. \ref{fig2}), viz., the Cu1-Cu2 bond length, decreases progressively as the structure 
changes from the tetragonal to the monoclinic phase. Consequently, we find an increase in the band gap with a reduction in the Cu1–Cu2 separation, akin to the Peierls-like distortion \cite{Peierls1955} that favors insulating behavior. Also, the Cu-F bond lengths evolve with distortion. For example, in the tetragonal structure, two of the Cu-F bonds were equivalent, while a difference in bond lengths appears with distortion as depicted in Fig. \ref{fig5}b. 

\begin{figure}[!t]
    \centering
\includegraphics[width=0.44\textwidth]{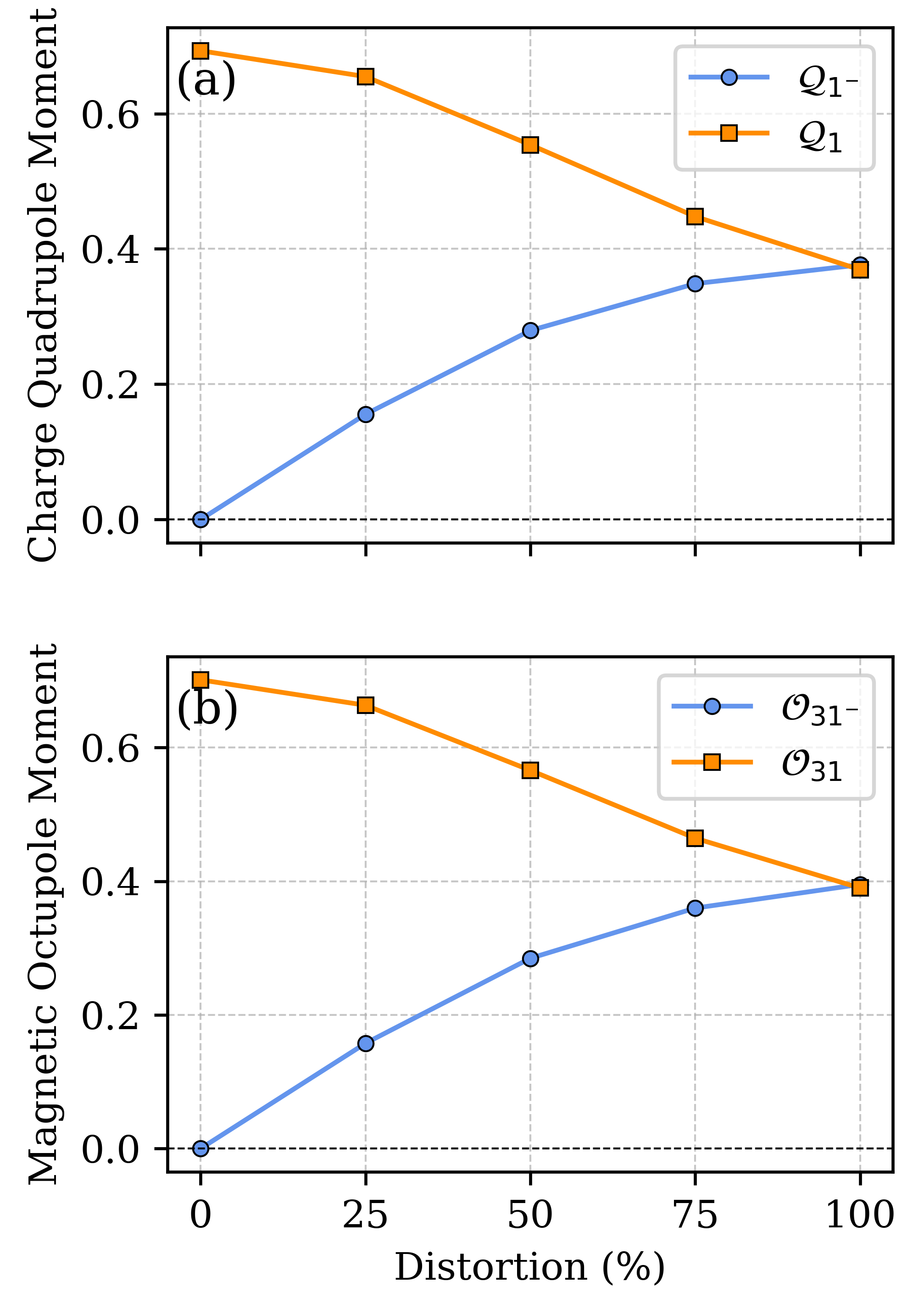}
    \caption{(a) Variation in the ferroically ordered magnetic octupole components ${\cal O}_{31}$ and  ${\cal O}_{{31}^-}$ (in $\mu_B$) as a function of amplitude of distortion. (b) Variation in the charge quadrupole components ${\cal Q}_{1}$ and  ${\cal Q}_{1^-}$ (in units of electronic charge) as a function of distortion amplitude. }
    \label{fig6}
    \vspace{-0.5em}
\end{figure}

We then investigate the variation of multipoles and the corresponding NRSS across the structural transition. In particular, we focus on the variation of magnetic octupoles and charge quadrupoles. 
The key results of our computed charge quadrupole and magnetic octupoles across the structural transition are shown in Fig. \ref{fig6}a and b, respectively. As seen from the plots, and from our earlier discussions in section \ref{Multipole Analysis}, there is only one antiferroic charge quadrupole component \(\mathcal{Q}_{{1}}\) and a ferroic magnetic octupole component \(\mathcal{O}_{{31}}\) present in the tetragonal structure. With the increase in distortion, both \(\mathcal{Q}_{{1}}\) and \(\mathcal{O}_{{31}}\) gradually decrease in magnitude. In contrast, distortion induces an additional antiferroic charge quadrupole component \(\mathcal{Q}_{{1}^{-}}\) and consequently a ferroic magnetic octupole component \(\mathcal{O}_{{31}^{-}}\), both of which are absent in the tetragonal structure. With an increase in distortion, the magnitudes of both these multipoles also increase, as evident from Fig. \ref{fig6}.

\begin{figure}[!t]
    \includegraphics[width=0.47\textwidth]{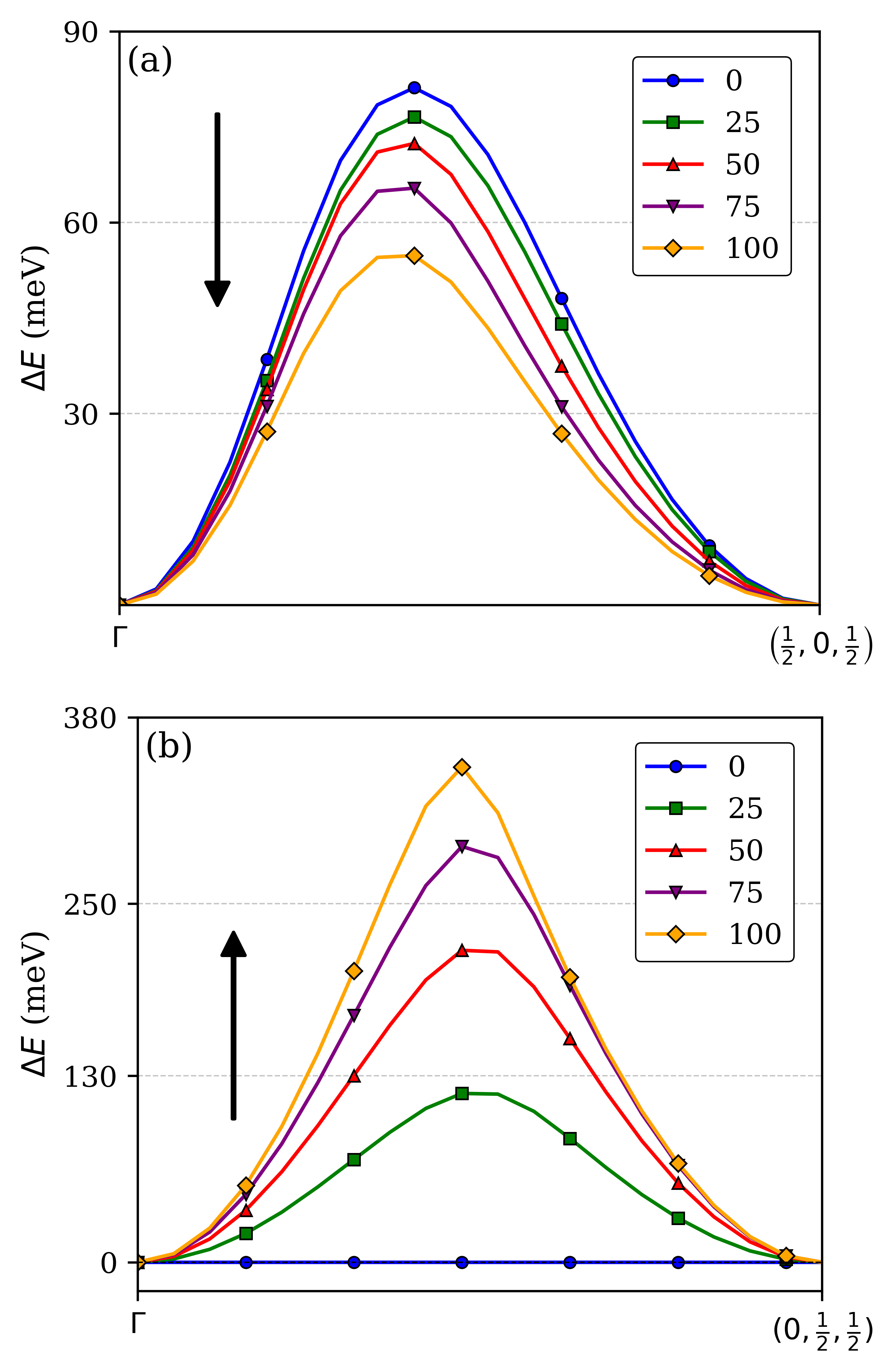}
    \caption{NRSS along (a) \(k_x\)–\(k_z\) and (b) \(k_y\)–\(k_z\) directions with varying amplitude of distortion. The direction of the arrow represents the increasing amplitude of distortion.}
    \label{fig7}
    \vspace{-0.5em}
\end{figure}

We next analyze the corresponding changes in the band structure, in particular, the NRSS energy \(\Delta E = E_{\uparrow} - E_{\downarrow}\), where \(E_{\uparrow}\) and \(E_{\downarrow}\) denote the energies of the spin-up and spin-down bands, respectively. Fig. \ref{fig7}a and b depict the computed changes in $\Delta E$ along the \(k_x\)–\(k_z\) and \(k_y\)–\(k_z\) directions of the BZ as a function of increasing amplitude of distortion. To compute $\Delta E$ along the \(k_x\)–\(k_z\), we consider the third pair of valence bands from the top, as this pair of bands remains energetically well-isolated from the neighboring bands along the \(k_x\)–\(k_z\) direction,  making it suitable for reliable comparison of $\Delta E$ across the different structures. As seen from Fig. \ref{fig7}a, $\Delta E$ along the \(k_x\)–\(k_z\) decreases in magnitude with increase in the amplitude of distortion, consistent with the corresponding changes in the ferroic magnetic octupole component \(\mathcal{O}_{{31}} \). We note that this is also consistent with our previous argument that the octupole component \(\mathcal{O}_{{31}}\) is associated with the NRSS along the [101] direction in reciprocal space.

In contrast, the spin splitting energy $\Delta E$ between the topmost pair of valence bands, that are well-separated from the rest along the \(k_y\)–\(k_z\) direction, is absent along \(k_y\)–\(k_z\) direction of the BZ of the tetragonal CuF$_2$ with 0\% distortion. With the increase in the amplitude of distortion, however, it becomes non-zero and also increases in magnitude (see Fig. \ref{fig7}b). We note that the change in $\Delta E$ along \(k_y\)–\(k_z\) direction is consistent with the increasing magnitude of the ferroic magnetic octupole component \(\mathcal{O}_{{31}^{-}}\), the $k$-space representation of which correlates to the NRSS along \(k_y\)–\(k_z\) direction. 

We note that the emergence of the NRSS, additionally along the \(k_y\)–\(k_z\) direction with the introduction of distortion to the tetragonal structure, gives rise to bulk-like $ d$-wave splitting. Interestingly, we find that our results in Figs. \ref{fig6} and \ref{fig7} remain qualitatively the same even for the second path along which the $\Gamma_5^+$ mode dominates. This highlights the importance of the $\Gamma_5^+$ distortion mode, describing the anti-polar motion of the F ions in giving rise to bulk-$d$ wave splitting in CuF$_2$ in contrast to the other members of the family that crystallize in the tetragonal structure and exhibit planar $d$ wave splitting.

\begin{figure}[!t]
    \centering
    \includegraphics[width=0.5\textwidth]{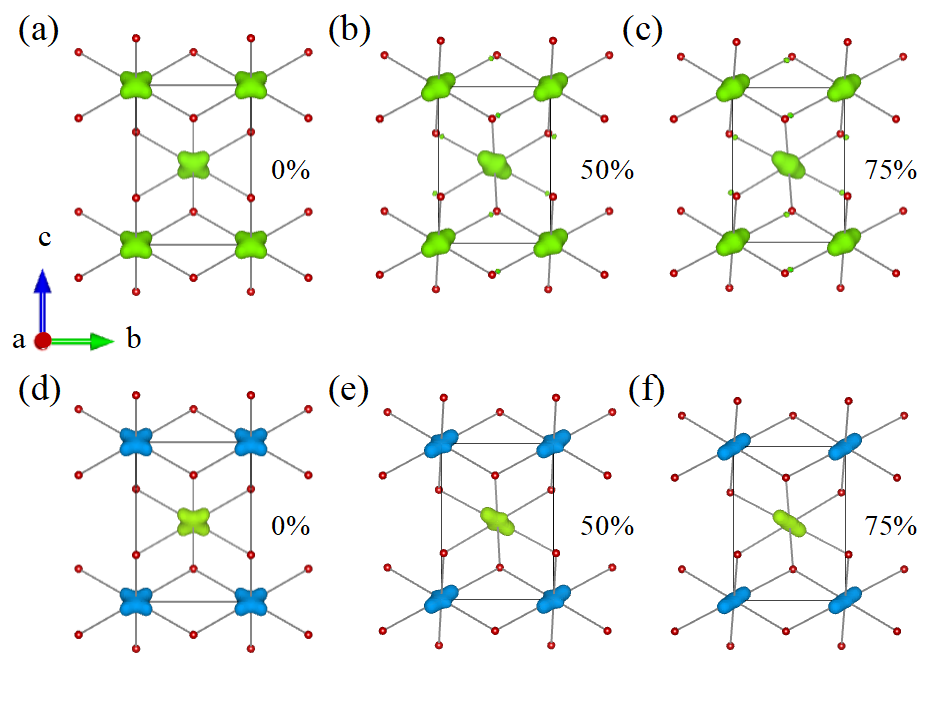}
    \caption{Band-decomposed charge density on the $bc$ plane of CuF$_2$ in the presence of (a) 0\% (tetragonal structure), (b) 50 \%, and (c) 75 \% amplitude of distortion. (d)-(f) shows the same for the band-decomposed magnetization density.}
    \label{fig8}
    \vspace{-0.5em}
\end{figure}

In order to further understand the effect of the phonon distortion modes on the NRSS physically, we compute and analyze the variation in charge and magnetization density across different intermediate structures. Fig. \ref{fig8}a-c depict the band decomposed charge density with increasing magnitude of distortion on the $bc$ plane of CuF$_2$. As evident from the figures, the charge density on the $bc$ plane is symmetric for the tetragonal structure with 0\% distortion, while it gradually becomes anisotropic with the increase in distortion. This, consequently, explains the emergence of the charge quadrupole component ${\cal Q}_{1^-}$ with increasing amplitude of distortion. We further investigate the corresponding changes in the magnetization density. As seen from Figs. \ref{fig8} d-f, we find that the magnetization density follows the charge density distribution, and accordingly, the magnetization density also gradually becomes more and more anisotropic on the $bc$ plane with increasing amplitude of distortion. Since the \(\mathcal{O}_{{31}^{-}}\) component quantifies the anisotropic magnetization density on the $yz$ plane, consequently we find a gradual increase in the \(\mathcal{O}_{{31}^{-}}\) component, leading to the emergence of NRSS along the \( k_y-k_z \) direction in the momentum space for the monoclinic structure.

\subsection{Effect of spin-orbit coupling}

The spin splitting in CuF$_2$ has so far discussed in the absence of SOC. In real materials, SOC is, however, present, and, therefore, we explicitly investigate the effect of SOC on the computed magnetic multipoles and the band structure of CuF$_2$.  

We begin by revisiting the electronic structure of CuF$_2$ in the presence of SOC. We analyze the symmetry-allowed magnetic configurations of the monoclinic structure of CuF$_2$ for the propagation vector $\vec k = (0,0,0)$ using the MAXMAGN utility of the Bilbao Crystallographic Server~\cite{Aroyo2006}. We find that the symmetry allows for only two magnetic space groups, (a) $P2_1'/c'$ and (b) $P2_1/c$.
Both magnetic space groups describe canted AFM configurations. The magnetic space group $P2_1'/c'$ corresponds to an AFM ordering of the $m_z$ component of the  Cu1 and Cu2 magnetic moments, while their $m_x$ and $m_y$ components are parallel to each other. 
On the other hand, the magnetic space group $P2_1/c$ depicts a canted AFM configuration with antiparallel $m_x$ and $m_y$ magnetic components at the Cu1 and Cu2 ions, and a parallel $m_z$ component. 

To determine the magnetic ground state of CuF$_2$ in the presence of SOC, we compute the total energies for both magnetic configurations, and the results of our calculations are presented in Table~\ref{tab3}. As evident from Table \ref{tab3}, the $P2_1'/c'$ configuration is slightly lower in energy than the $P2_1/c$ configuration, suggesting a canted AFM ground state with a tiny net magnetic moment (see Table \ref{tab3}) resulting from the spin canting.   

\begin{table}[htbp]
\centering
\caption{Comparison of total energies and spin moment at Cu ions for the magnetic configurations, corresponding to the two symmetry-allowed magnetic space groups (see text for details). The energy difference $\Delta E$ is set to zero for the lowest energy configuration.} 
\begin{tabular}{|l|c|c|c|}
\hline
Magnetic & $\Delta E$ & Magnetic moment & Total \\
Spacegroup & meV & [$\mu_B/$Cu] & Moment [$\mu_B$] \\
\hline
 $P2_1'/c'$ & 0  & 0.78 & 0.08 \\
 $P2_1/c$   & 0.2 & 0.77 & 0.07 \\
\hline
\end{tabular}
\label{tab3}
\end{table}

\begin{figure}[!t]
    \centering
\includegraphics[width=0.45\textwidth]{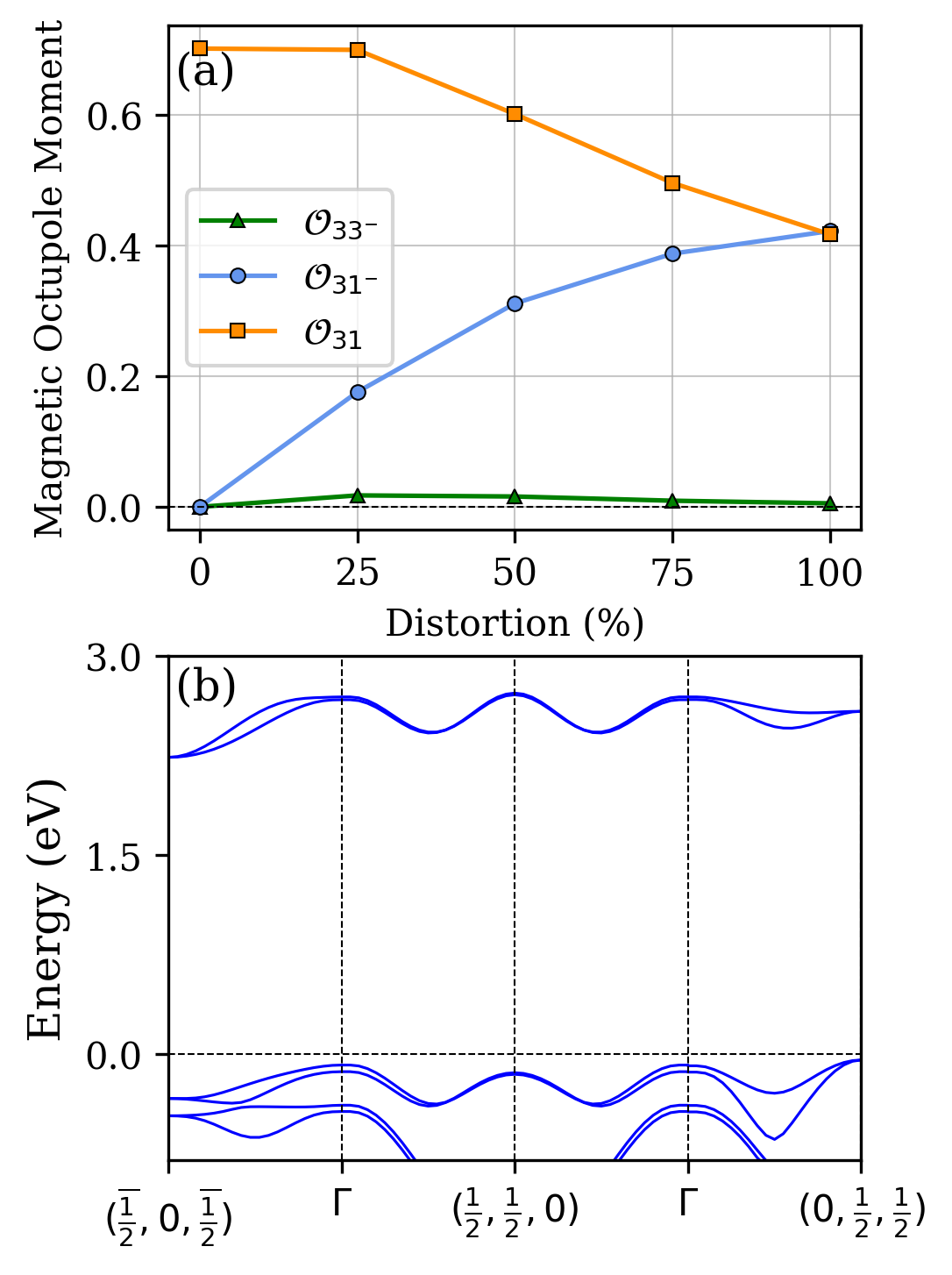}
    \caption{Effect of SOC. (a) Variation of the ferroically ordered magnetic octupole components as a function of distortion in the presence of spin-orbit coupling. (b) Band structure of \(\mathrm{CuF_2} \) in the presence of SOC. }
    \label{fig9}
    \vspace{-0.5em} 
\end{figure}

We then compute the atomic-site magnetic multipoles for the energetically lower $P2_1'/c'$ magnetic configuration in the presence of SOC. Our calculations show that in addition to the existing components \(\mathcal{O}_{31}\) and \(\mathcal{O}_{31^-}\) in the absence of SOC, a new ferroically ordered octupole component, \(\mathcal{O}_{33^-}\), also becomes non-zero. The existence of the ferroically ordered octupole component \(\mathcal{O}_{33^-}\) is also consistent with the magnetic space group  $P2_1'/c'$. We further investigate the effect of distortion on the ferroically ordered octupole components in the presence of SOC, and the results are shown in Fig. \ref{fig9}a. As seen from the figure, the presence of SOC does not affect the qualitative dependence of the octupole components \(\mathcal{O}_{31}\) and \(\mathcal{O}_{31^-}\) on the structural distortion in the absence of SOC, as discussed in section \ref{Structural distortion}. In contrast, the new SOC-induced component \(\mathcal{O}_{33^-}\) exhibits only a weak dependence on the structural distortion and has a much smaller magnitude compared to the other two components, as shown in Fig.~\ref{fig9}.

We further investigate the effect of SOC on the band structure of CuF$_2$ in the $P2_1'/c'$ magnetic configuration. Our computed band structure (see Fig. \ref{fig9}b) shows that the NRSS along the \(k_y\)–\(k_z\) and \(k_x\)–\(k_z\) directions remains unaffected by the SOC. However, the tiny net magnetic moment, resulting from the SOC-induced spin canting, leads to a spin splitting of the bands even at the \(\Gamma\)-point of the BZ, as evident from Fig.~\ref{fig9}b. 
Furthermore, referring back to the \(k\)-space representations of magnetic octupole moments, the emergence of the octupole component \(\mathcal{O}_{33^-}\)
suggests the presence of spin splitting along the additional \( k_x \)-\( k_y \)  direction. Indeed, our computed band structure, as depicted in Fig.~\ref{fig9}b, confirms the presence of a small but non-negligible splitting along the \( k_x \)-\( k_y \)  direction due to the inclusion of SOC, consistent with the presence of \(\mathcal{O}_{{33}^{-}}\) multipole.

\section{Summary and Outlook} \label{sec4}

To summarize, we have developed a multipolar framework to understand the origin of the bulk $d$-wave NRSS in monoclinic CuF$_2$. While many members of the $M$F$_2$ family, such as MnF$_2$ and CoF$_2$, crystallize in the tetragonal $P4_2/mnm$ rutile structure, this structure is not observed for CuF$_2$, likely due to the smaller ionic radius of Cu$^{2+}$. Consistently, our phonon calculations reveal an instability of the $P4_2/mnm$ structure in CuF$_2$, following which, we identify antipolar F displacements that stabilize the low-symmetry monoclinic $P2_1/c$ structure. By systematically comparing the ground-state $P2_1/c$ structure with the hypothetical $P4_2/mnm$ tetragonal structure, we demonstrate that these antipolar F displacements play a decisive role in generating additional ferroically ordered magnetic octupoles that transform the planar $d$-wave NRSS, characteristic of rutile-type MF$_2$ compounds, into a bulk-like $d$-wave pattern.

Our multipolar analysis reveals that the tetragonal structure hosts only one totally symmetric component of the ferroic magnetic octupole, resulting in planar NRSS similar to that observed in MnF$_2$. In contrast, the monoclinic symmetry of CuF$_2$ gives rise to an additional antiferroic charge quadrupole component and, consequently, a second totally symmetric magnetic octupole component. Their corresponding $k$-space representations accordingly lead to the emergence of NRSS along distinct momentum directions, consistent with the transition from planar to bulk $d$-wave splitting.

It is important to point out that even though $P4_2/mnm$ is a hypothetical structure of CuF$_2$, the $P4_2/mnm \rightarrow P2_1/c$ transformation can be observed in CoF$_2$ under 45 GPa of pressure \cite{Cihan_CoF2_2016}. At an intermediate pressure, around 6.5–8 GPa, the orthorhombic $Pnnm$ phase is also found \cite{Cihan_CoF2_2016, Barreda_CoF2_2013} in CoF$_2$, which is linked to the $\Gamma_2^+$ atomic distortion, as discussed in the present work. Thus, the role of structural distortion in understanding the differences found in the NRSS pattern may be crucial and practically realizable through pressure-induced structural transitions.

Another important finding of our work is the role of SOC in the electronic structure. Even though NRSS does not require the presence of SOC, SOC is always present in real materials. Our detailed investigation shows that while SOC does not modify the essential NRSS features, it leads to additional interesting features in the band structure. For example, in the case of CuF$_2$, we find that SOC induces weak spin canting that generates a small net magnetic moment and an additional ferroically ordered octupole component, leading to additional splittings, most notably along the $k_x$–$k_y$ plane, and also lifting the degeneracy of bands. It is important to note that SOC-induced additional spin splitting may contribute significantly to anomalous Hall transport properties in metallic systems, highlighting the importance of the interplay between SOC-induced relativistic splitting and NRSS.

Our work highlights the crucial role of structural distortions and multipolar degrees of freedom in stabilizing bulk $d$-wave NRSS in CuF$_2$, and provides a route to engineering the momentum-space pattern of NRSS in centrosymmetric antiferromagnets without changing the N\'eel order using external perturbations such as strain or pressure. These findings complement ongoing efforts to control the magnitude and sign of NRSS, motivating further exploration along these directions.

\section{Acknowledgment}
M and SBh thank the National Supercomputing Mission for providing computing resources
of ‘PARAM Porul’ at NIT Trichy, implemented by C-DAC and supported by the Ministry
of Electronics and Information Technology (MeitY) and Department of Science and
Technology, Government of India. SBa acknowledges HPC resources and support provided by CINECA under the ISCRA IsCc9 IDMVW project. 
SBh gratefully acknowledges financial support from the IRCC Seed Grant (Project Code: RD/0523-IRCCSH0-018), the INSPIRE Research Grant (Project Code: RD/0124-DST0030-002), and the ANRF PMECRG Grant (Project Code: RD/0125-ANRF000-019).

\appendix

\begin{figure}[t]
\includegraphics[width=0.45\textwidth]{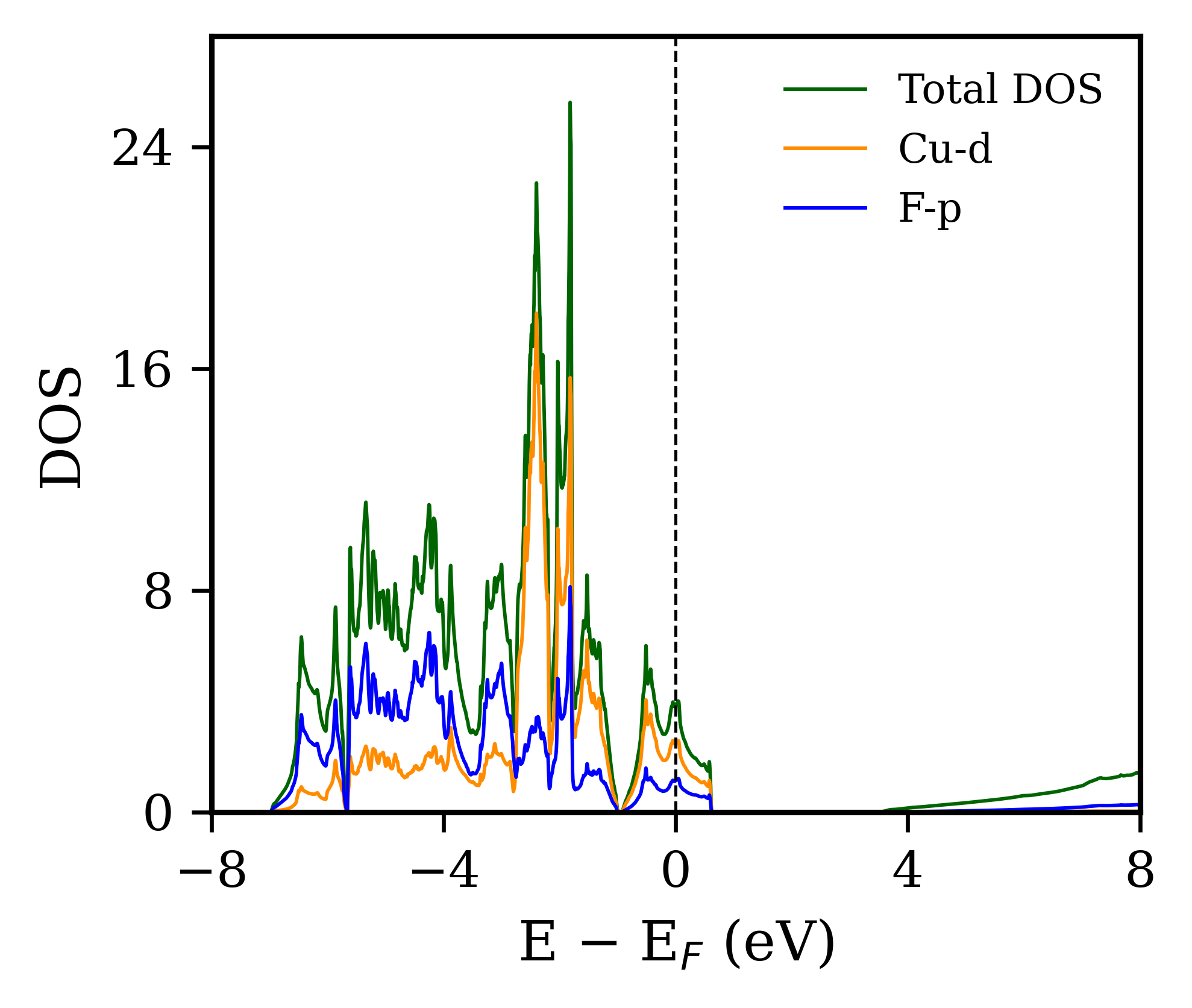} 
   \caption{Density of states (DOS) for CuF\textsubscript{2}, computed within the non-magnetic configuration. The total DOS and the partial DOS for the Cu--\emph{d} and F--\emph{p} orbitals are shown. The Fermi energy is set to zero.}
    \label{fig10}
    \vspace{-0.5em} 
\end{figure}

\section{Nonmagnetic densities of states}\label{app1}

We have computed the electronic structure of CuF\textsubscript{2} within the non-magnetic configuration. The resulting density of states (DOS) is shown in Fig. \ref{fig10}.  As seen from the plot, there is a
non-zero DOS at the Fermi energy, indicating metallic behavior, consistent with the corresponding band structure in Fig. \ref{fig3}a. The partial DOS, as shown in Fig. \ref{fig10}, suggests that the states across the Fermi energy are primarily derived from the Cu -\emph{d} orbitals. The hybridization between Cu-\emph{d} and
F-\emph{p} orbitals, which extend across the Fermi energy, is also evident from the plot.

\bibliography{main}

\end{document}